# Changing Identities and Evolving Conceptions of Inquiry through Teacher-Driven Professional Development

**Ben Van Dusen, Mike Ross, and Valerie Otero, CU Boulder**

### Abstract

This STEM education study investigates the "Streamline to Mastery" professional development program, in which teachers work in partnership with university researchers to design professional development opportunities for themselves and for fellow teachers. Our research describes the process of teacher professional growth both through changes in agency and through a shared pursuit of an improved understanding of classroom scientific inquiry. Videos, emails, lesson reflections, survey responses, and interviews were analyzed to glean insight into changes in teachers' discourse around inquiry and into their shifts in participation within the professional community they established. Implications for professional development in STEM education are discussed.

Our country is in dire need of highly-qualified physics teachers (NRC, 2007). A majority of those who teach physics are teaching outside of their content areas, with only 35% holding a major in physics or physics education (Hodapp, Hehn & Hein, 2009; NRC, 2007). Recruitment is an important priority, but so too is professional development for the 65% of teachers who are teaching physics outside their areas of expertise.

Streamline to Mastery is a five-year professional development program for teachers who find themselves teaching physics and physical science. The teachers are charged with partnering with university researchers to establish a community committed to improving science education. Because teachers must play an active and central role in the design of the program, it is derived from the expertise and experiences of the teachers who participate. Streamline to Mastery follows an experiential learning model where a high level of trust and a small amount of guidance are provided with the intent of cultivating an environment of openness, skepticism, and candor that results in the generation of knowledge that is relevant to its participants.

The goals of Streamline to Mastery are to support teachers in improving their professional practices and to develop a community of science education leaders within the greater population of practicing teachers. These are the only formalized goals of this professional development program; more specific goals must, by design, emerge from the teachers' own perceived needs and areas of interest. This model is based on the perspective of *communities of practice* (Grossman, Wineburg, & Woolworth, 2010; Lave & Wenger, 1991; Wenger, 1998), which defines learning in terms of participation within a community of likeminded scholars. The idea is that when a group of expert learners collaborates around a shared inquiry, knowledge is generated through skeptical discourse, individual agency, and shared experiences and prior knowledge. In developing the Streamline to Mastery program, we hypothesized that if we could provide a context that would encourage teachers to articulate their thoughts about physics teaching and learning, challenge the ideas of other teachers, and critically analyze data from their own classrooms, they would establish a sense of agency, leadership, and enhanced knowledge regarding teaching physics.

In order to ensure that this professional development program is driven by the needs of the teachers it serves, the teacher participants' concerns and perceived areas for growth were elicited early in the first year. The first Streamline cohort chose to focus on the topic of inquiry-oriented science instruction, and this has been a dominant thread in the first eighteen months of the professional development program. As science education researchers have noted, a shared understanding of the term *inquiry* has been elusive. It is common for both researchers and practitioners to characterize inquiry teaching and learning in disparate ways (Anderson, 2002; Windschitl, 2004). Researchers have investigated pre-service teachers' conceptions of inquiry and found them to be inconsistent with those of practicing scientists, and others have found experienced science teachers' conceptions of classroom scientific inquiry to be incomplete as compared to NRC published documents (Kang & et al., 2008) such as *Inquiry and the NSES* (NRC, 2000).

This study attempts to shed light on how a small sample of practicing science teachers created a shared understanding of classroom scientific inquiry while receiving no explicit instruction on this or any other related topic. This study also examines how these teachers created knowledge, such as a more sophisticated understanding of inquiry, by engaging in self-driven reflective practice and community discourse. We address the following questions: (1) *How did the participating teachers' talk about inquiry teaching and learning change throughout the course of this study?* and (2) *In what ways did the nature of Streamline teachers' participation and leadership roles change in their interactions with other teachers and researchers?*

## Research Context

Four teachers from two urban high schools in the Rocky Mountain West were recruited as the first of two cohorts of secondary physical science teachers to participate in Streamline to Mastery. The teachers have between two and seven years of classroom experience and three of the four teach outside of their scientific discipline, as shown in Table 1. Each of the teacher participants work in schools with a majority of students living in poverty and a majority of students from racial or ethnic groups underrepresented in STEM careers.

Requirements to be in the Streamline program include teaching in a high needs district, completion of a master's degree, and a willingness to share and collaborate on aspects of one's teaching practice. Teachers and researchers met every other week to share lessons, plan research, and discuss topics of interest to the teachers. The *lesson-sharing,* in which teachers and researchers each shared a lesson that they deemed to be effective and inquiry-oriented with the other teachers and researchers, was a one of the central activities during the first several months of the program. The teacher participants also attended numerous national conferences, including one in which they presented a poster on the Streamline to Mastery program and one in which they collectively presented a workshop on inquiry-oriented science instruction.

TABLE 1. Participant Demographics

| Degree | Years Exp. | Subject Taught |
|---|---|---|
| B.A.Bio/Ph.D. Biochem | 2 | Physics |
| B.A. Chem/M.A. Urban Ed | 3 | Physics |
| B.A. Phys/M.A. Urban Ed | 3 | Physical Sci. |
| B.A. Bio/M.A. Urban Ed | 7 | Physical Sci. |

**Theoretical Framework**

This study draws from a perspective on learning that emphasizes the critical role of social interaction in the development of individual mental processes (Michaels, O'Connor, & Resnick, 2008; Vygotsky, 1986; Wertsch, 1991). Thus, learning, viewed here as neither solely individual nor completely communal, occurs as actors participate in communities of practice, often toward a common goal. Such goal-directed activities were present throughout much of the Streamline community's discourse, and our shared pursuit of greater understanding of our own classroom practices served as motivation for the productive talk that led to the creation of a shared understanding of ideas that were relevant, both individually and collectively, to the participants.

Academically productive talk has been shown to support learning and enhanced understanding (Michaels, O'Connor, & Resnick, 2008), and we espouse a view that learning occurs *through talk* as persons participate in communities of practice (Lave & Wenger, 1991). That is, through talk, actors learn the practices and forms of discourse that are valued by and define a community of practice. Accountability to the norms and values of a learning community, as well as accountability to standards of reason and knowledge characterize talk that supports learning and understanding. It is through this social engagement that actors may co-construct community norms and practices, as well as form their own identities in relation to the community. Thus, we take the nature of these communities, and their practices and forms of discourse, to be dynamic systems of social relations.

Another point of emphasis concerning the forms of productive talk in which our community engaged is that of talk rooted in participants' experiences. Our view of the importance of relevant experiences is consistent with those of Dewey (1998), as well as with Vygotsksy's emphasis on the importance of concrete, or everyday, experiences in the learning process present in his theory of concept formation (1986). Specifically, that everyday experience and formalized academic concepts mediate one another as one learns. Academically productive talk, we assert, can be a highly constructive mechanism for such mediation. Participants drawing from common experiences may lead to collectively value forms of practice and discourse and this process can be a means for the creation of new knowledge that is relevant to their community. Such a lens has proven useful in our attempts to understand the forms of discourse that emerged from and ultimately came to define the Streamline to Mastery's community of practice, as well as the changes in collective agency and identity that were evident in this community.

**Methods**

<u>Teacher Talk about Inquiry</u>

The data collected for this portion of the study concerning teachers' talk about inquiry teaching and learning consists of lesson sharing reflections, responses to two administrations of prompts taken from a survey developed to measure teachers' conceptions of inquiry, and video of professional development meetings. These data were used to triangulate our claims about teachers' talk about inquiry and were analyzed using the five essential features of inquiry specified in the NRC document *Inquiry and the National Science Education Standards: A guide for teaching and learning* (2000).

These five essential features and their hereafter abbreviated names are (1) Engaging in scientifically oriented questions (Questions), (2) Giving priority to evidence (Evidence), (3) Formulating explanations based on evidence (Explain), (4) Evaluating explanations in connection with scientific knowledge (Connections), and (5) Communicating explanations (Communicate). Though this research team does not assert that the NSES description of inquiry should or does represent a "gold standard" for inquiry-oriented instruction, we chose to employ it as a framework for assessing teachers' talk about inquiry and consider talk consistent with this framework to be "expert-like" for the purposes of this study. It should be noted that the researchers never made this framework or other related literature available to the teacher participants.

### Inquiry Survey Item Responses

The inquiry survey used in this study was designed by Wallace and Kang (2004) to assess secondary science teachers' conceptions of inquiry relative to the five essential features of inquiry. As an example, the scenario "Giving students a white powder" was designed specifically to elicit a response related to the inquiry feature *Giving priority to evidence* (Evidence). Rather than using the survey items as the designers did, we used the items to cue extended open-ended responses about the topic of inquiry more generally (all items are show in Appendix A). These response data were coded for the five essential features as well as other notable response patterns not captured in the NSES inquiry framework.

### Lesson-sharing Reflections

In addition to survey item responses, the participants also generated reflections following each of the five lesson sharing events. As stated above, the teachers and researchers each chose a lesson that they deemed to be effective and inquiry-oriented to teach to the group. After the lesson sharing, teachers and researchers debriefed aspects of the lessons together and teachers completed one online lesson-sharing reflection for each lesson. The five lesson-sharing reflections were responses to the prompts: (1) *In what ways was this an inquiry lesson?* and (2) *How might you modify this lesson for your classroom?* These were recorded using an online message board and participants could see the posts made by the others.

The five lesson-sharing reflections occurred in the first five months of the study, and, because inquiry was a recurring topic of teacher discussion and concern throughout the first year, the inquiry survey items were administered in the 10[th] month of the project to assess teacher talk about inquiry. The survey items were administered again in the 17[th] month of the project to gain more longitudinal data. The lesson sharing and survey timeline is shown in Figure 1.

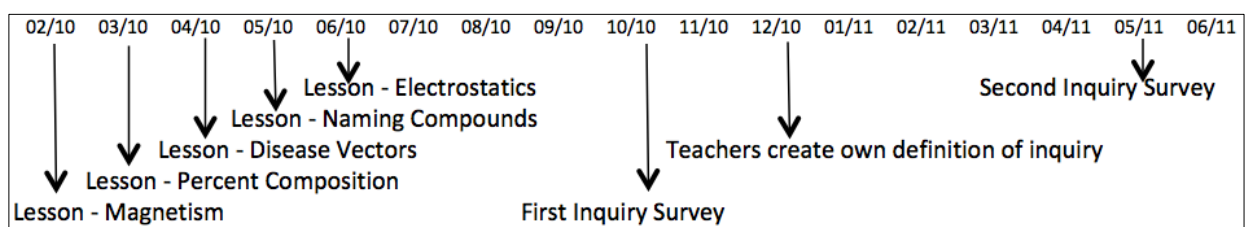

**Figure 1.** Timeline of Lesson-Sharing and Survey Responses

### Teacher Agency

In order to investigate the second research question involving the growth of teachers' participation and leadership, we coded videotapes of the meetings, analyzed email threads, and interviewed the teachers. We collected forty-two hours of video. Initially, the videotapes were coded using a generative coding scheme to determine dominant themes in the meetings. For example, we found *agenda setting* to be a useful code for analyzing teachers' changes in participation over time. In addition, email threads were analyzed to determine any increases in numbers or types of emails sent by the teachers. To address research question two, we focused on a code labeled *challenging*, which marks when a teacher or researcher challenges the statement of another group member, to evaluate both the nature of the discourse and whether it reflected what could be considered a community of practice.

In the first phase of the analysis, we examined each video to determine who had set the agendas. To do so, each segment of text in which a new agenda item was begun was coded according to whether a teacher or researcher introduced the agenda item. Agenda items were considered primary if they were related to the overarching structure of the day, such as lesson sharing, and coded secondary if they came along during conversation. As an example of a secondary instance, a teacher might say, "Let's not forget to talk about the workshop." One point was given to primary agenda items and 0.5 points were given to secondary. Examples of typical agenda items include: discussing research data, planning for workshops, and planning for next year's meetings.

As a secondary method for analyzing the video, the research team generatively developed a coding scheme to illustrate the substance of the semi-weekly meetings. The coding scheme was created through researchers individually coding the transcripts and comparing their codes to the rest of the research team. This iterative process led to the creation of our complete set of codes, which have been used to inform two research projects. We arrived at twelve codes, including *challenging*, *conversation driving*, and *vulnerability*. For the purpose of this analysis, we only report on the *challenging* code. The researchers came to a consensus on all of the codes on a small subset of the transcripts; however, we are still in the process of testing for more general inter-rater reliability.

The group's emails were used as a second data source to analyze teacher participation and roles within the group. These emails were analyzed to determine from whom they were sent (teacher or researcher), the date of their origination, what their primary topic was, and if they represented a new line of discussion or were a reply to another email. Example discussion topics included scheduling meetings and offering resources to the group.

We triangulated our findings using a third source of data from a focus group interview that one researcher conducted with the teachers. The teachers were asked to recall how their roles and preparation for meetings had changed over time. The teachers were also asked to explain their process of preparing for two presentations they made at the Western Regional NOYCE Conference. Finally, a member check was performed.

## Findings

Teacher Talk about Inquiry

At the start of the study, teachers' lesson sharing reflections and meeting talk were ambiguous with regard to the subject of inquiry. The term *inquiry* was frequently used by teachers in a manner that made it difficult for the researchers to distinguish its intended meaning from other terms used frequently such as *constructivism, hands-on, real-world,* and even *best practice* based on the context of the conversations and reflections. For example, when responding to the lesson-sharing reflection prompt asking what made the share lessons inquiry-oriented, teachers offered the responses shown in Table 2.

Table 2. Teacher responses to inquiry oriented lesson sharing

| Teacher | Statement |
|---------|-----------|
| 1 | …it really focused on kids trying to figure it out for themselves. A true hands-on activity. |
| 2 | As a result of the process and labs, students construct an understanding of how carriers… |
| 3 | This [name of curriculum] really does guide students' thinking through an abstract concept starting with a real-life application… |

Teacher 2 used the terms *inquiry* and *constructivism* interchangeably in meeting conversations: *"How can I use already created materials while upholding a constructivism or inquiry approach in the classroom?"* Teacher 4 used terminology ambiguously as well, but offered expert-like responses to the lesson-sharing prompt following the first lesson share: *"Students were really doing science by testing their model and revising it based on experimental evidence".*

As shown in Figure 2, the frequency with which the teacher participants characterized the shared lessons' inquiry-orientation using the five essential features of inquiry increased over the course of the five lessons. In lessons 1 and 3, only g*iving priority to evidence* (Evidence) and *formulating explanations based on evidence* (Explain) were noted. In lesson 2, no references were made that aligned with the five essential features of inquiry. By the fifth lesson, however, four of the five essential features were referenced. Only e*valuating explanations in connection with scientific knowledge* (Connect) was not referenced. These data show increasing numbers of references consistent with the NSES inquiry framework over the course of the five lessons. It should also be noted that the coded references were relatively evenly distributed across the four participants over each of the data sources.

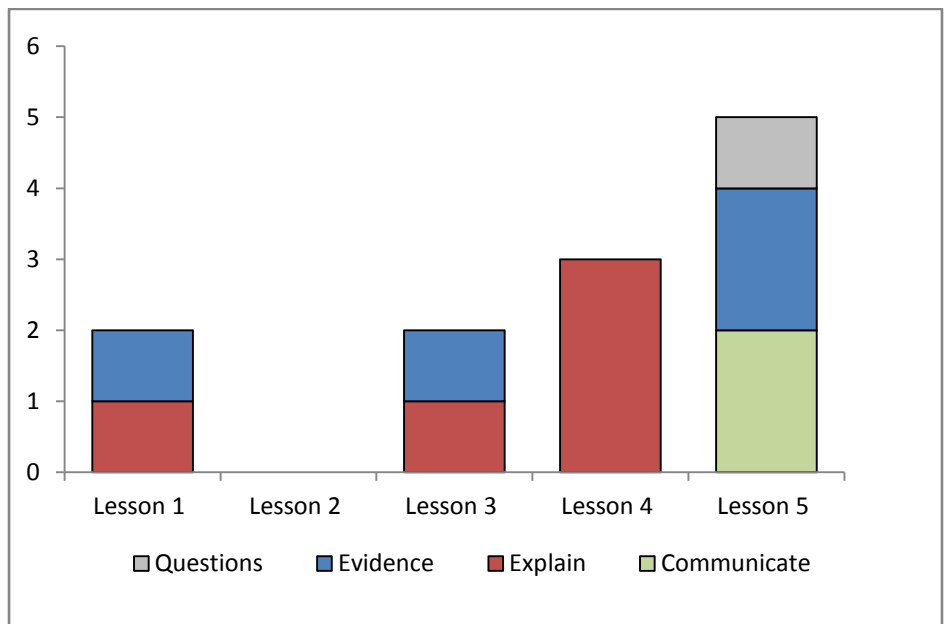

Figure 2. Frequency of References to Five Essential Features of Inquiry in Lesson Reflections

References not captured by the five NSES essential features of inquiry were observed and coded, as they were deemed relevant to characterizing these teachers' talk about inquiry and often appeared frequently in other data sources. As shown in Figure 3, references to the Real-world oriented tasks (Real-world) were common as were references to the social nature of student activities (Social), ownership of ideas, tasks, procedures, etc. (Own), scientific models (Model), and constructivist epistemology (Constructivism).

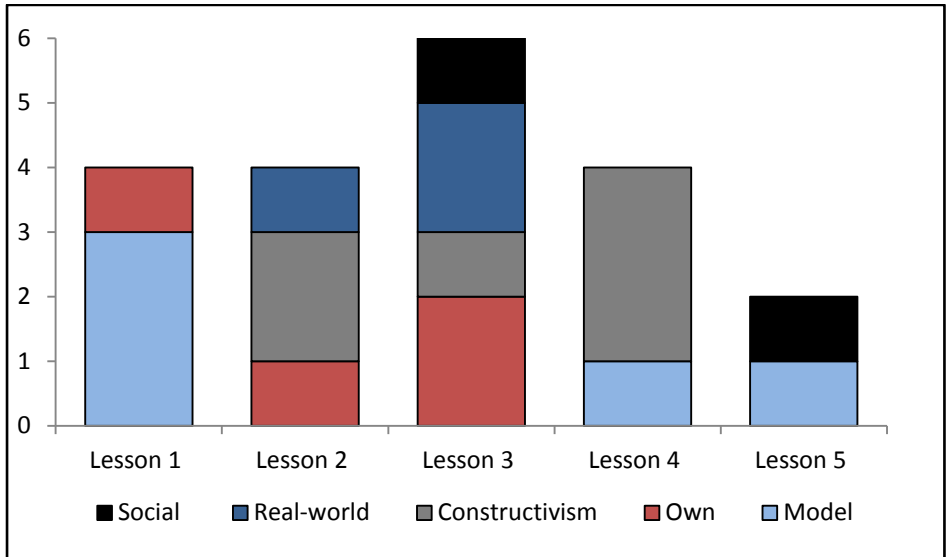

Figure 3. Frequency of References Not Captured in the NSES Inquiry Framework

The inquiry survey data were coded using the same system developed for the lesson-sharing reflections and, though the survey was identical in the two administrations, the results bore some notable differences. As shown in Figure 4, the frequency with which teachers referenced *Communicate* and *Explain* increased approximately five-fold for each category, and the frequency with which they referenced *Evidence* decreased from 13 to 11. As with the lesson-sharing reflection data, no references to *Connect* were made. The increase in the number of references that were consistent with the NSES inquiry framework is further evidence of these teachers' evolution toward a more expert-like understanding of inquiry.

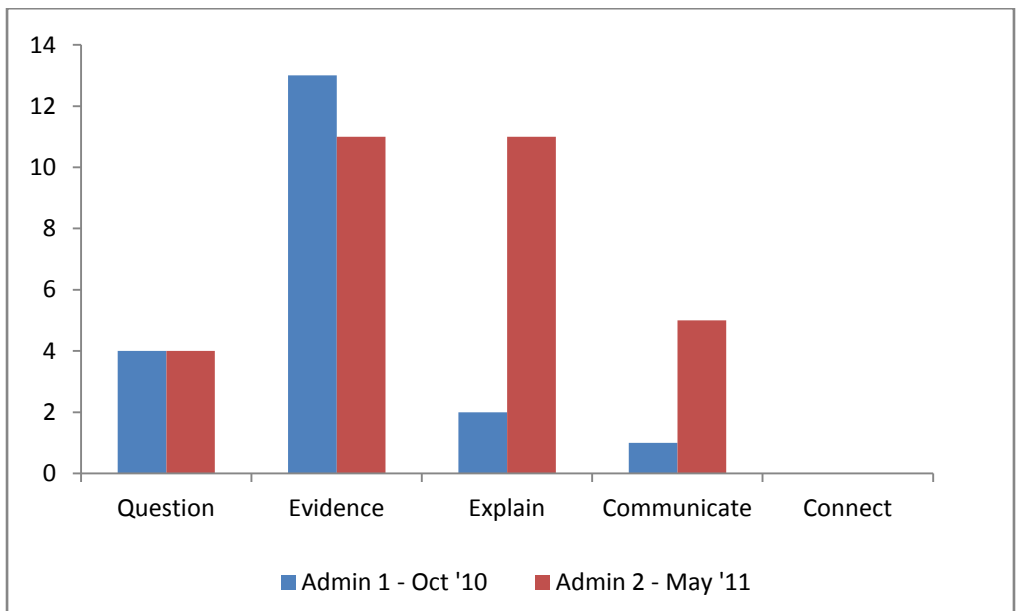

Figure 4. Frequency of References to Five Essential Features on Inquiry Survey Items

As is shown in Figure 5, references to *Real-world* and *Model* decreased markedly from the first to second administrations, and references to *Social* decreased by half. References to *Constructivism* and *Own* increased moderately. Though some variation is noted across the two administrations, it is clear from the data that these teachers feel that ownership of ideas, social construction of knowledge, and real-world relevance are also key features of inquiry.

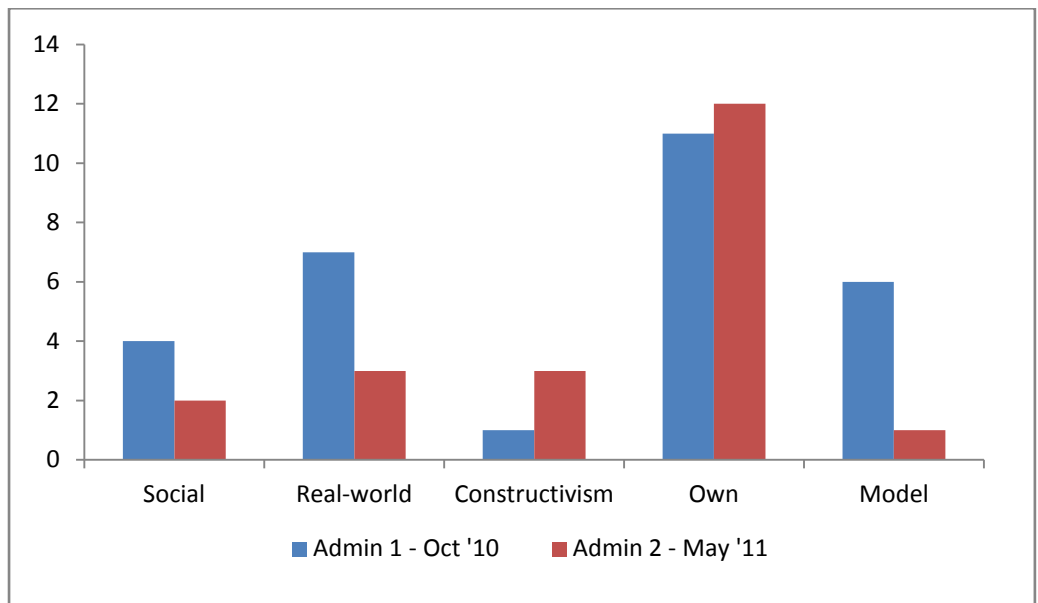

Figure 5. Frequency of References Not Captured in the NSES Inquiry Framework

Finally, it is important to note that in the 12[th] month of the study, Teacher 2 motivated a discussion with the aim of defining inquiry. Teacher 2 had attended a conference and engaged with colleagues in a conversation about inquiry learning. Upon returning Teacher 2 shares:*"I felt like I should have a more cohesive ability to discuss it, or more cohesive description of inquiry, with everything that we've done, and I felt like I was, like, somewhat articulate, but not as much as I should be."* This began a conversation in which a shared meaning of inquiry was established. Through a 1.5 hour conversation driven by the teacher participants with minimal facilitation by the researchers, the teachers arrived at their own definition of inquiry: *"Socially constructing evidence-based meaning of phenomena through intentionally sequenced events."*

Throughout this study, we have documented evidence of the evolution of our teacher participants' talk about inquiry. The teachers appeared to progress through five stages, represented in Figure 6. We claim that these teachers' reached some shared understanding around the topic of inquiry, though we acknowledge that their trajectories toward this understanding, as well as their current conceptions of classroom scientific inquiry, likely exhibit some variation. As shown in Figure 6, the Streamline teachers began the study talking ambiguously about inquiry (Phase I), but through our common activities and talk came to realize that their own ideas on the topic were unclear (Phase II). This realization motivated a collective development of a shared definition of inquiry (Phase III).

Presentations made at national conferences by the Streamline to Mastery teachers were evidence of teachers entering Phase IV, *Coherent and Consistent in Dialogue.* That is, these teachers represented a collective and shared understanding of inquiry at two conferences during this study and one more recently in late 2011. Coherence and consistent implementation of inquiry in practice, Phase V, however, is meant to represent what our research team envisions as a goal of this program, and is as yet empirically unverified.

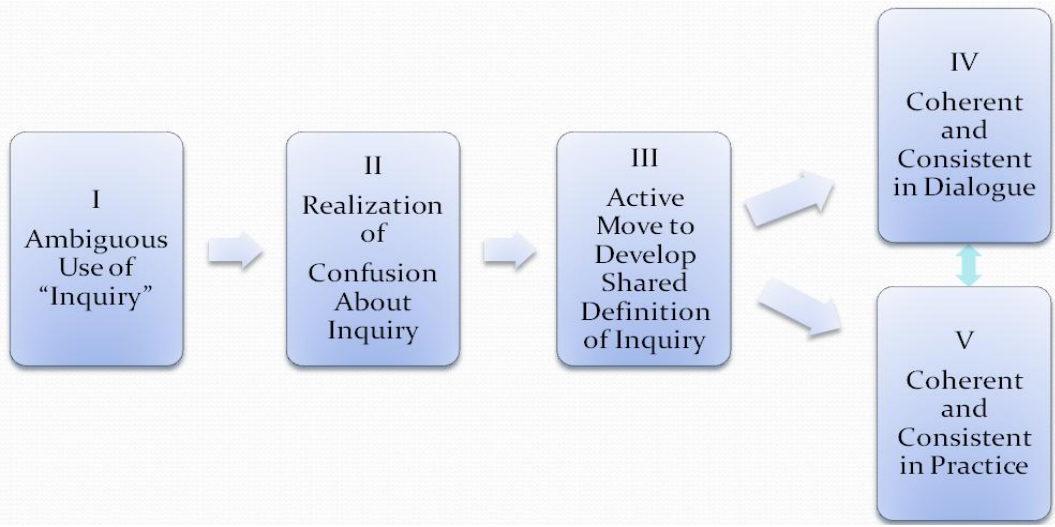

Figure 6. Phases of Participating Teachers' Conceptions of Inquiry.

<u>Teacher Agency</u>

Analysis of emails yielded the following results. Initially, the researchers sent the majority of the emails, but over time the teachers began to send a larger share of the total number of emails. With the exception of the final month, the percentage of the emails originating from teachers increased. These results are shown in Figure 7. The total number of emails in a given month ranged from eight to fifty-four. During the earlier period, designated by section (a) in Figure 7, the majority of the emails were from researchers and focused on scheduling meetings with the occasional email about meeting agenda items.

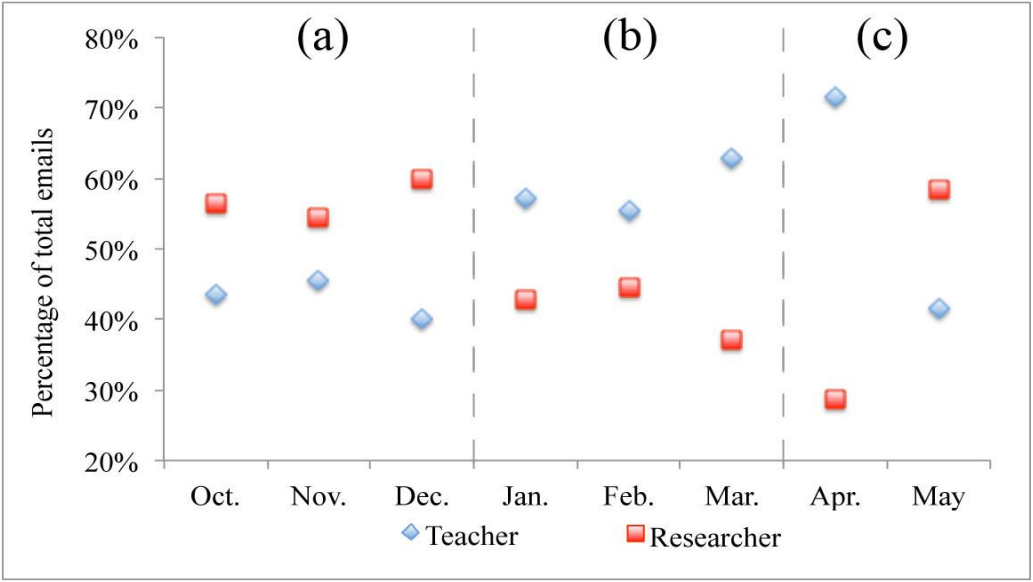

Figure 7. Percentage of total emails sent each month.

During the middle period, designated by section (b), teachers began to write more of the emails, in which they are primarily either asking for assistance or acting as a resource for one another. The March emails are dominated by the teachers' preparation for presenting a poster and holding a workshop at the Western Regional NOYCE Conference. During the late period, designated by section (c), we see teachers sending an increased percentage of the emails in April when they were sharing resources and scheduling meetings after their workshop. The month of May shows a resurgence of emails originating from researchers as they gave feedback to the teachers on their PERC papers.

We also analyzed the same set of emails to determine whether the teachers or researchers were the ones beginning the *conversation threads*. In this analysis a slightly different pattern emerges. Figure 8 shows the percentage of new email conversation threads by month, again broken down into early, middle, and late time periods. During the early time period (a) researchers began almost all of the new conversations. In the middle time period (b) conversations originating from teachers and researchers were evenly balanced with the exception of March. Again, during March the teachers were preparing several presentations for the Western Regional NOYCE Conference, which required significant coordination among the teachers. During this time the researchers primarily acted as resources in answering teacher questions. During the late time period (c) the number of email conversations begun by teachers and researchers was balanced.

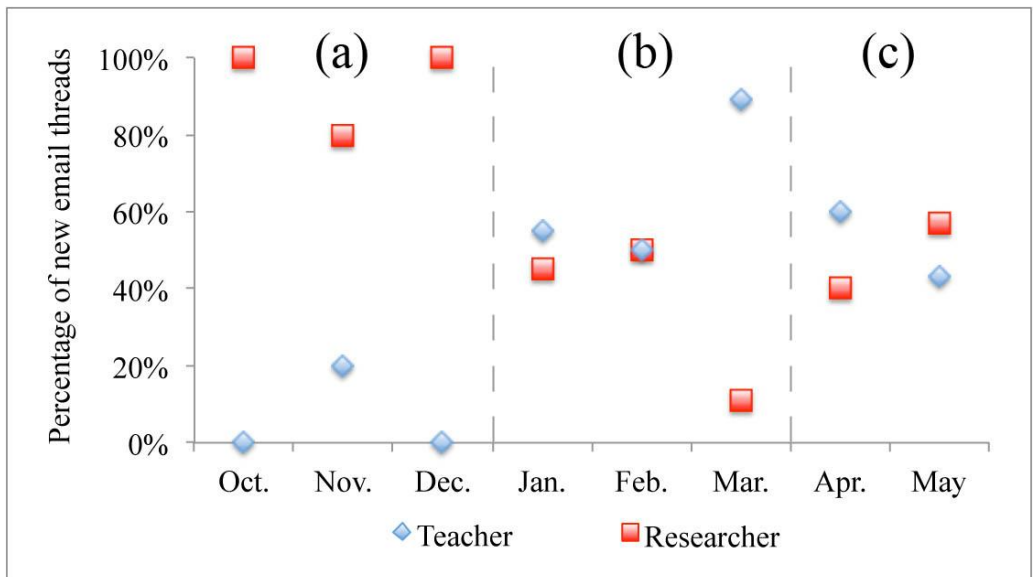

Figure 8. Percentage of new email discussion threads.

From the analysis of email activity, we see an increasing share of email activity from teachers, shown in Figure 7, which suggests that teacher participation increased over time. We also see the increasing share of new email threads from teachers, shown in Figure 8, which suggests that the teachers increased their involvement and initiative for leadership. Inferences about these findings will be made in the next section.

We now turn to findings of the video data analysis of the meetings in which we coded teachers' and researchers' contributions to meeting agenda items and conversation driving. To illustrate the change in agenda setting, we averaged data within each time period (early, middle, and late). As shown in Figure 9(a), researchers provided all of the meeting agenda items during the early time period. The middle time period shows substantial growth in the percentage of agenda items provided by teachers. During the late section (c), the teachers provided the majority of the agenda items. The change that takes place between (a) and (b) came largely from the teachers beginning to take charge of the meetings, ultimately taking on a majority of the agenda setting responsibility.

During a focus group interview, the teachers confirmed that the teachers are aware of the changes in the community that are shown in Figures 7, 8, and 9. Teachers stated their awareness of their new roles as agenda setters and conversation drivers, "I feel in the beginning a lot of it was kind of done for us, like we just showed up… I've noticed that Valerie (Otero) started to step out of that role of prepping the agenda and we've stepped up into that."

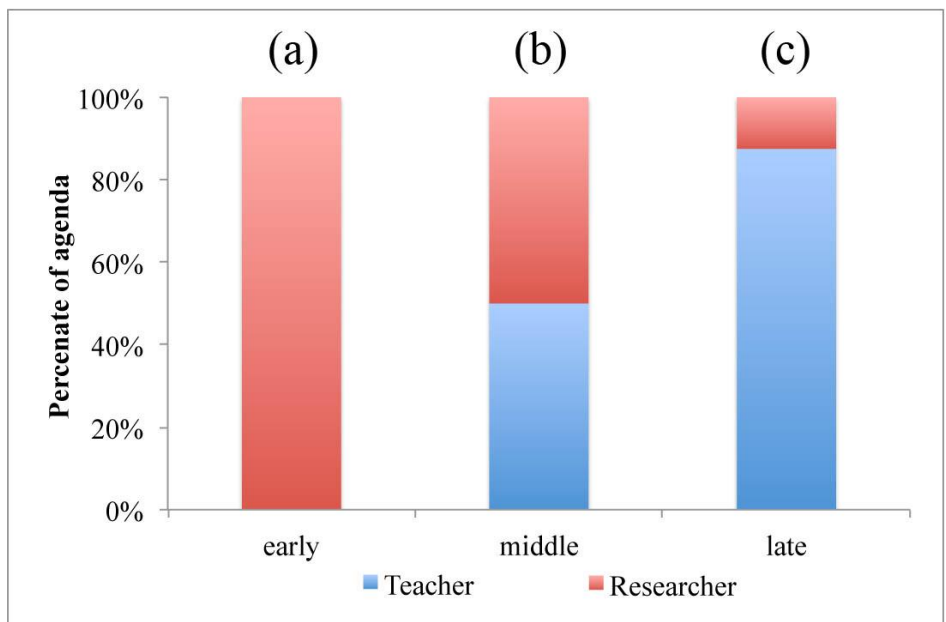

Figure 9. Agenda items from teachers and researchers.

Focusing on the challenging code in our video analysis, we found that in our early meetings neither the teachers nor researchers challenged each others' statements. Responses tended to be supportive and reassuring. During the middle time period, however, we see over a four-fold increase in the number of challenging statements. Challenging was often done explicitly, for example, T1: "Can I play devil's [advocate]?" T2: "Yeah, devil away!" By the end of the year, teachers were engaging in debates ranging from disagreements about responsibilities as Streamline teachers to whether activities were inquiry-based.

In this section of the paper we reported on four main findings of the Streamline community. First, teachers increasingly participated in the email discourse over time. Second, teachers started to begin email threads on their own. Third, teachers began to set agendas. Finally, teachers became increasingly comfortable challenging each others' ideas.

## Conclusion

We claim that the Streamline community learned, as is evident by their co-construction of meaning of the concept of inquiry and in the changing participation of its members over time. As the teachers engaged in collaborative discourse—discourse in which drew directly from their classroom experiences—their understanding appeared to evolve. Through this process, the teachers came to the realization that their own ideas about inquiry had been unclear and in need of refinement. Though we cannot account for other maturation effects, we attribute the evolution of the community's talk about and understanding of inquiry to the academically productive talk that occurred. We hypothesize that a key feature of this talk is the sustained and explicit reflection on and grounding in the lesson-sharing and classroom experiences of the participants.

We also observed two forms of changing participation. First, we saw an increase in total teacher participation as shown by increased emails as well as an increase in leadership roles as shown by increased email thread origination, conversation driving, and agenda setting. Second, we saw a shifting of roles from a hierarchical community in which the researchers were the experts and the teachers were the learners, to a more egalitarian community where everyone participates as expert learners. In a community of practice where everyone is an expert learner, there must be constant willingness to share ideas and to challenge one another's ideas, as well as an acceptance of, and affinity for, skepticism in order for growth to take place. And indeed, this is what we observed. Over time, teachers increasingly shared their ideas and challenged one another. Our evidence supports the claim that a community of practice is forming among these teachers (Grossman, 2001; Wenger, 1998) in which vulnerability and skepticism allows for the growth and development of the community as well as the growth of its individual members.

These findings raise important questions concerning the preparation of teachers for the physics classroom. How can we expect teachers to understand inquiry teaching and learning when the current national efforts to enhance inquiry learning in science appear to have failed? First, when we think about teacher change we might stop thinking about how to *make* people change according to broad and general assertions about what improvements are needed in K-12 science schooling, and instead think about how to *create communities* in which change might happen. Such change, we insist, must be relevant to the particularities of teachers' practices and contexts. Second, when we think about professional development programs we often think about bringing expertise and resources *to* the teachers. We might instead explore the view that the resources necessary for meaningful professional development reside *within* the teachers and their particular contexts. By leveraging a community's everyday professional experiences and insights to address their shared goals, we can create substantive opportunity for professional growth.

**Appendix A** – Inquiry Survey Items

1. Having students gather data for a local nonprofit organization.
2. Giving students a white powder and asking them to determine what the powder is.
3. Asking students to develop and answer their own questions about a local wetlands area.
4. Having students follow a procedure to complete a lab.
5. Asking students to use what they know about a local forest to decide whether an old-folks home should be built on that land.
6. Having students classify substances based upon their observable properties.
7. Having students use graphics on the Internet to explain how gas molecules move.
8. Having students make presentations of data collected during a lab.
9. Asking students to improve on a basic design (make an airplane fly further, make a motor spin faster, etc.).
10. A class discussion about the arrangement of the periodic table.